\documentclass[12pt]{article}
		\usepackage{epsfig, amssymb}
		\usepackage{graphicx,psfrag} 
		\usepackage{mathrsfs}
		\usepackage{euscript}
		\usepackage{bbold}
		\usepackage{amsmath}
		\usepackage{stmaryrd}
		\usepackage{wasysym}%
		\usepackage[hang,flushmargin]{footmisc} 
		\renewcommand{\baselinestretch}{1.4}
		\DeclareMathAlphabet{\mathpzc}{OT1}{pzc}{m}{it}

		\usepackage[ps2pdf,bookmarks=true,colorlinks,linkcolor=red,urlcolor=blue,citecolor=blue]{hyperref}
		\usepackage[usenames]{color}
		\newcommand\CL{{\mathcal L}}
		\newcommand\CQ{{\mathcal Q}}

		\newcommand\CM{{\mathcal M}}

		\newcommand\CO{\mathcal{O}}

		\newcommand\CN{{\mathcal N}}

		\newcommand\chif{\tilde{\chi}}

		\def\CA{{\cal A}}

		\def\CM{{\cal M}}
		\def\CN{{\cal N}}
		\def\CO{{\cal O}}
		
		\def\CQ{{\cal Q}}
		\def\CK{{\cal K}}
		
		\def\CL{{\cal L}}

		\def\II{\relax{I\kern-.10em I}}

		\def\IZ{\relax{\rm Z\kern-.34em Z}}
		\def\IB{\relax{\rm I\kern-.18em B}}
		\def\IC{{\relax\hbox{$\inbar\kern-.3em{\rm C}$}}}
		\def\ID{\relax{\rm I\kern-.18em D}}
		\def\IE{\relax{\rm I\kern-.18em E}}
		\def\IF{\relax{\rm I\kern-.18em F}}
		\def\IG{\relax\hbox{$\inbar\kern-.3em{\rm G}$}}
		\def\IGa{\relax\hbox{${\rm I}\kern-.18em\Gamma$}}
		\def\IH{\relax{\rm I\kern-.18em H}}
		\def\II{\relax{\rm I\kern-.18em I}}
		\def\IK{\relax{\rm I\kern-.18em K}}
		\def\IP{\relax{\rm I\kern-.18em P}}
		
		\def\inbar{\,\vrule height1.5ex width.4pt depth0pt}

		\def\IR{\relax{\rm I\kern-.18em R}}

		\def\lp10{\ell_p^{10}}
		\def\lp11{\ell_p^{11}}
		\def\R11{R_{11}}

		\def\frac#1#2{{#1 \over #2}}

		\newcommand{\be}{\begin{equation}}
		\newcommand{\ee}{\end{equation}}

		\newcommand{ \jsn}{\mathfrak{sn}}
		\newcommand{ \jcn}{\mathfrak{cn}}
		\newcommand{ \jdn}{\mathfrak{dn}}
		\newcommand{\jam}{\mathfrak{am}}

		\newcommand{\LPR}{\cite{Luty:2012ww}}

		\newdimen\tableauside\tableauside=1.0ex
		\newdimen\tableaurule\tableaurule=0.4pt
		\newdimen\tableaustep
		\def\phantomhrule#1{\hbox{\vbox to0pt{\hrule height\tableaurule width#1\vss}}}
		\def\phantomvrule#1{\vbox{\hbox to0pt{\vrule width\tableaurule height#1\hss}}}
		\def\sqr{\vbox{%
		  \phantomhrule\tableaustep
		  \hbox{\phantomvrule\tableaustep\kern\tableaustep\phantomvrule\tableaustep}%
		  \hbox{\vbox{\phantomhrule\tableauside}\kern-\tableaurule}}}
		\def\squares#1{\hbox{\count0=#1\noindent\loop\sqr
		  \advance\count0 by-1 \ifnum\count0>0\repeat}}
		\def\tableau#1{\vcenter{\offinterlineskip
		  \tableaustep=\tableauside\advance\tableaustep by-\tableaurule
		  \kern\normallineskip\hbox
		    {\kern\normallineskip\vbox
		      {\gettableau#1 0 }%
		     \kern\normallineskip\kern\tableaurule}%
		  \kern\normallineskip\kern\tableaurule}}
		\def\gettableau#1 {\ifnum#1=0\let\next=\null\else
		  \squares{#1}\let\next=\gettableau\fi\next}
		
		 \def\eqnn#1{\xdef #1{(\secsym\the\meqno)}\writedef{#1\leftbracket#1}%
		 \global\advance\meqno by1\wrlabeL#1}
		 \def\eqna#1{\xdef #1##1{\hbox{$(\secsym\the\meqno##1)$}}
		 \writedef{#1\numbersign1\leftbracket#1{\numbersign1}}%
		 \global\advance\meqno by1\wrlabeL{#1$\{\}$}}
		 \def\eqn#1#2{\xdef #1{(\secsym\the\meqno)}\writedef{#1\leftbracket#1}%
		 \global\advance\meqno by1$$#2\eqno#1\eqlabeL#1$$}

		\global\newcount\itemno \global\itemno=0
		
		\def\itemaut#1{\global\advance\itemno by1\noindent\item{\the\itemno.}#1}
		
		\def\del{\partial}
		
		\def\({\left(}
		\def\){\right)}
		
		\def\eg{{\it e.g.}}

		\def\etal{{\it et.~al.}}
		\newif{\ifeq}           
		\eqtrue                 
		\numberwithin{equation}{section}
		
\begin{document}

		\begin{titlepage}
		
		\begin{flushright}
		YITP-SB-12-44\\
		MIT-CTP/4396
		\end{flushright}
		\vskip.2in
		\begin{center}
		{\huge Gravity duals of cyclic RG flows, with strings attached}\\
		\end{center}
		\vskip.3in
		\begin{center}
		{\large Koushik Balasubramanian}\\
		\vspace{0.15cm}
		C. N. Yang Institute for Theoretical Physics, Stony Brook University, Stony Brook, NY 11764, USA\\
		\vspace{0.25cm}
		Center for Theoretical Physics, MIT, Cambridge, MA 02139, USA\\
		\end{center}
		\vspace{0.25cm}
		\begin{center}
		{\large ABSTRACT}
		\end{center}
		{
		In this note, we propose gravity duals for 3+1 dimensional Lorentz invariant theories exhibiting discrete scale invariance. We construct non-singular solutions of a six dimensional gravitational theory that are warped products of $AdS_{5}$ and a circle. The presence of non-trivial warp factor explicitly breaks the symmetries of $AdS_{5}$ to discrete scale invariance and also breaks translation symmetry along the circle. The matter content of the 6D gravitational theory {\it does not violate} the null energy condition. In addition, we show that the linearized fluctuations around these 6D backgrounds are stable {\it i.e.,} the fluctuations do not violate the Breitenlohner-Freedman stability criterion. The dual theories display periodic RG trajectories, but these RG flows do not violate the $a$-theorem. In particular, the dual field theory is {\it not a soft deformation }(deformation by marginal or relevant operators) of a 3+1 dimensional conformal field theory and hence the $a$-theorem and similar monotonicity constraints on RG flows do not apply to these examples. The holographic $c$-theorem also does not apply to these solutions. Finally, we present solutions of type II supergravity that exhibit discrete scale invariance to demonstrate that such solutions can be embedded in string theory.
		 }

		\noindent
		\vspace{3mm}

		\end{titlepage}
		\newpage
		\renewcommand{\baselinestretch}{1.1}  
		\setcounter{page}{1}
		
		\newpage
		
		\section{Introduction} 
		\noindent
		
		Most applications of renormalization group equations involve a flow towards a fixed point. In general, the RG flows can end in more general attractors such as, limit cycles or even chaotic trajectories.  In  \cite{Wilson:1970ag}, Wilson pointed out that phenomena exhibiting log-periodic behavior in momentum scale are described by RG limit cycles. Such theories are invariant under a discrete scale transformation. Many examples of {\it non-relativistic} unitary theories exhibiting discrete scale invariance (DSI) have been found \cite{discrete}. Such theories can be helpful in understanding interesting universal phenomena arising in condensed matter system (for instance, Efimov states). 
		 It would be interesting to know what kind of theories can exhibit this behavior.
		
		The possibility of such exotic RG flows in a unitary Lorentz invariant theory is somewhat restricted due to certain monotonicity constraints.
		In the case of 1+1 dimensional unitary, renormalizable quantum field theories (QFT), Zamolodchikov \cite{Zamolodchikov:1986gt} proved the existence of a function (of couplings) $C$ that decreases monotonically along the RG flow. The function $C$ becomes stationary at fixed points and its value at a fixed point is given by the central charge of the conformal field theory (CFT) describing the fixed point. By completing an argument of Zamolodchikov, Polchinski \cite{Polchinski:1987dy} showed that a 2D renormalizable QFT (unitary and Lorentz invariant) exhibiting continuous scale invariance is also conformally invariant. Since then there have  been many developments in establishing such monotoncity results in other dimensions. 
		
		Recently, Komargadski and Schwimmer \cite{Komargodski:2011vj} proved the $a$-theorem which establishes the monotonicity of the $a$ function along RG flows between fixed points in 3+1 dimensions. The $a$ function interpolates between the Euler anomaly coefficients of the UV ($a_{UV}$) and IR ($a_{IR}$) fixed points. Using the analysis of \cite{Komargodski:2011vj}, Luty {\it et al.} \LPR\, showed that  perturbative scale invariant theories in 3+1 dimensions are also conformally invariant. 
		They argued that RG flows in 3+1 dimensions can only asymptote to fixed points or limit cycles that are equivalent to fixed points.{\footnote{We will refer to such limit cycles as conformal limit cycles. These limit cycles are associated with conformal invariance and hence different from limit cycles associated with discrete scale invariance.}}
		Some counterexamples (which do not violate the assumptions in \LPR) of this result were suggested in \cite{GrinsteinPast1, GrinsteinPast2}. 
			However, it was pointed out in a later paper \cite{Luty:2012ww, Fortin:2012hc, Fortin:2012hn} that the counterexamples proposed in \cite{GrinsteinPast2} are actually conformally invariant. In particular, they point out that the examples in \cite{GrinsteinPast1, GrinsteinPast2} are conformally invariant even though the $\beta-$functions do not vanish (based on earlier results by Jack and Osborn \cite{JackOsborn, Osborn}). 

		These constraints on RG flows seem to rule out the possibility of periodic or chaotic RG flows in unitary Lorentz invariant theories.\footnote{{Such flows might exist when the flow function is multivalued in the space of couplings \cite{Curtright:2011qg}.}} 
				However, these results apply to flows starting from a UV theory that can be described as marginal or relevant deformations (soft deformations) of a conformal field theory (CFT) in 3+1 dimensions. The monotonicity results of \cite{Komargodski:2011vj} and \LPR\, in general, do not apply to flows where the UV description contains non-renormalizable interactions. In fact, Anselmi {\it et al.} \cite{Anselmi:1997ys} showed that $a_{IR}$ is not less than $a_{UV}$ for certain RG flows triggered by non-renormalizable deformations. However, the interpretation of $a_{UV}$ for such an RG flow is ambiguous as the flow does not interpolate between undeformed ``UV" CFT and the IR CFT. 
These RG flows were studied earlier in \cite{Kutasov:1995np, Kutasov:1995ss} where they argue that the non-renormalizable interactions are {\it dangerously irrelevant}. 
				
				There are examples of interacting non-renormalizable theories that are scale invariant (continuous) but not conformal \cite{Jackiw} in 3+1 dimensions (also see \LPR\, for a discussion). However, the dynamics of these theories is not well understood due to their unconventional kinetic terms \cite{Jackiw}. It is also worth noting that in an effective field theory description of Efimov physics, it is essential to include irrelevant deformations to get rid of divergences in certain three body scattering amplitudes. 
		It seems worthwhile to look for Lorentz invariant theories exhibiting DSI that cannot be described as soft deformations of a CFT. 
		
		A non-renormalizable UV theory can be interpreted as an effective description for energies $E$ less than some UV scale $\Lambda$.  However, to have a well-defined theory for all energies (including energies greater than $\Lambda$) the theory needs to be UV completed.
		In an effective field theory (EFT) framework, the deep IR behavior of a theory can be analyzed without worrying about the existence of a UV completion. The degrees of freedom with energies greater than the cut-off $\Lambda$ are ``integrated out" resulting in an effective theory with non-renormalizable interactions. It might appear that non-renormalizable interactions can be ignored in an effective theory as they become irrelevant in the deep IR. However, some non-renormalizable interactions could become relevant (or marginal) along the RG flow in the presence of other marginal or relevant deformations. It is dangerous to ignore such non-renormalizable interactions and they are known as dangerous irrelevant deformations (see appendix of \cite{Kutasov:1995ss} for a discussion of such operators). These operators are perturbatively irrelevant but their effect cannot be ignored in the infrared. In general, the presence of such deformations leads to breakdown of a perturbative expansion in the couplings and a non-perturbative framework might be necessary to understand these deformations. In principle, it is possible to avoid dangerous irrelevant operators by UV completing the theory. However, the existence of a UV completion may not always provide a convenient framework for studying the low energy physics. {{For instance, the UV completion may not have a Lagrangian or any conventional field theory description.}}		
		
		Gauge/gravity correspondence \cite{Maldacena:1997re,Gubser:1998bc, Witten:1998qj} offers a convenient framework to study non-perturbative effects in theories with holographic duals.
		 In this note, we use the holographic principle to demonstrate the existence of 3+1 dimensional Lorentz invariant effective theories that exhibit discrete scale invariance. We argue that these holographic theories are not soft deformations of a conformal field theory. 
		Monotonicity constraints on RG flows suggest that these theories cannot be UV completed by 3+1 dimensional CFTs. In other words, it appears that discrete scale invariance provides an IR-obstruction to UV completion by a 3+1 D Lorentz invariant field theory.
However, UV completion by higher dimensional theories (on $\displaystyle{\mathbf{R}^{3,1} \times {\cal M}_{ compact}}$), non-Lorentz invariant field theories or lattice theories is not ruled out. We expect that the holographic DSI theories constructed in this paper can be UV completed by higher dimensional theories or by world volume theories on D-branes. We will provide some weak evidence for this claim. 		
		
		We will now discuss a strategy to find a holographic description of theories exhibiting DSI, without violating the holographic $c-$theorem \cite{Alvarez:1998wr, Freedman:1999gp, Myers:2010tj}. We will begin this discussion with a brief overview of the holographic $c-$theorem. This theorem is a monotonicity constraint on classical domain wall solutions (with $d-$dimensional Lorentz invariance)
		of a $d+1$ dimensional gravitational theory described by the following action
		\be S= {1\over 2 \kappa_{d+1}^2}\int d^{d+1}x \sqrt g \(R - 2 \Lambda + {\CL}_{matter} \)  + S_{bdy} \label{DWaction}\ee
		where ${\CL}_{matter}$ is the Lagrangian describing the matter content of the theory. We shall assume that the action does not involve higher spin fields and respects diffeomorphism invariance. The most general line element with $d-$dimensional Lorentz invariance, that extremizes this action has the following form:
		\be ds^2 = e^{2 A(w)} \left[ -dt ^2 + d\vec{x}^2 \right] + dw^2 \label{DWmetric} \ee
		This line element describes the metric on $AdS_{d+1}$ when $A(w) = w/L_{AdS}$. Using Einstein's equations of motion we find
		\be (d-1)A''(w) =  \left(T^{t}_{t} - T^{w}_{w}\right) \label{DWequation}\ee
		where $T_{\mu \nu} $ is the stress tensor due to the matter fields. We further assume that the stress tensor satisfies the null energy condition (NEC) $T_{\mu \nu} \xi^{\mu} \xi^{\nu} \ge 0$ for every future pointing null-vector $\xi^{\mu}$. This condition is not derived from any fundamental principle, but physically reasonable matter is believed to satisfy this condition (classically). This condition is usually imposed to eliminate superluminal effects and other peculiar features in a classical gravitational theory. The null energy condition in conjunction with equation (\ref{DWequation}) leads to the following monotonicity condition on the warp factor: 
		$A''(w) \le 0$. We did not have to use the fact that the metric is asymptotically $AdS$ to obtain this monotonicity relation. 
		We can define a quantity $a(w)$ as follows
		\be a(w) = \({\pi^{d/2} \over \Gamma\(d/2\) \kappa_{d+1}^2 A'(w)^{d-1}}\) \ee
		The monotonicity condition on the warp factor can be written as $a'(w) \ge 0$ {\it i.e.,} $a(w)$ is monotonic along the holographic direction. 
		This condition known as the holographic $c-$theorem, can also be obtained from the Raychaudhuri equation applied to null congruences \cite{Alvarez:1998wr, Sahakian:1999bd}. For solutions that describe a flow between $AdS_{5}^{UV}$ and $AdS_5^{IR}$, this condition leads to the holographic version of the $a-$theorem in 3+1 dimensions. For the class of 3+1-dimensional CFTs dual to classical Einstein gravity, $a=c$ \cite{Henningson:1998gx} {\it i.e.}, the coefficient of Euler density is the same  as the coefficient of the square of Weyl tensor in the trace anomaly. There is a generaliztion of the holographic monotonicity condition for gravity duals of theories with $a\neq c$ as well \cite{Myers:2010xs}.

		We will now prove a corollary of the holographic $c-$theorem. We will show that (assuming the NEC is satisfied) a regular metric of the form in (\ref{DWmetric}) cannot display discrete dilatation invariance if it approaches $AdS$ as $w \rightarrow \infty$. An asymptotically $AdS$ metric of the form (\ref{DWmetric}) can exhibit DSI if and only if, $A(w)$ can be written as $$ A(w) = {w\over L_{AdS}} + f(w)$$ where, $f(w)$ is a bounded periodic function. The metric is not regular if $f(w)$ is unbounded. Since $f(w)$ is periodic, $f''(w)$ must take both positive and negative values. This implies that $A''(w)$ must take both positive and negative values for the metric to exhibit discrete scale invariance. This cannot happen unless NEC is violated. Hence, an asymptotically $AdS$ metric of the form (\ref{DWmetric}) exhibiting discrete scale invariance is either supported by matter that violates the null energy condition or it must be singular. We would also like to point out that solutions with singularities can be physically meaningful if the singularities can be resolved.
		
		
		
		It is possible to find holographic examples of cyclic RG flows by violating the null energy condition.  
		Using a toy model, Nakayama \cite{Nakayama:2011zw, Nakayama:2009qu} showed that violation of null-energy condition can lead to solutions of (\ref{DWaction}) with DSI. In this model, conformal invariance is broken by a complex vector field that condenses. At the critical point, the energy-momentum tensor vanishes and appears to saturate the null-energy condition. However, the stress tensor does not satisfy the null-energy condition for all configurations of the matter field. In fact, the null energy condition is violated by arbitrarily small fluctuations around the critical point. Violation of the NEC generically have been argued to lead to instabilities quite generically \cite{NECinstabiity} suggesting that the dual field theory description of this toy model (if it exists) is unlikely to be unitary. But, there are consistent frameworks which allow for localized violations of the null-energy condition. In string theory, orientifold planes (which are negative tension objects) can allow for localized violation of null-energy condition.  Quantum effects can allow for localized violations of null-energy condition, but the {quantized matter fields} should satisfy an average null energy condition. It would be interesting to find a generalization of the holographic $c-$theorem to include such effects.  
		

		We will now discuss an approach to finding gravity duals of cyclic RG flows without violating the null-energy condition. Consider domain wall solutions (of a $d+D+1$ dimensional gravitational theory) with $d$ dimensional Lorentz invariance described by the following line element
		\be ds^2 = e^{2 A(w, \theta_{k})} \left[ e^{2 w/L} \left( -dt^2 + d\vec{x}^2 \right)+ dw^2 \right] + e^{2B(w,\theta_{k})}g_{i j} d\theta^{i} d\theta^{j} + 2 \zeta_{i}(w,\theta_{k}) dw d\theta^{i} \label{DSansatz} \ee
		where $\theta^{i}$ denotes an internal direction and $i \in 1,...,D$. The above line element describes a warped product of $AdS_{d+1}$ and $D-$dimensional compact internal manifold $\CM_{D}$, $AdS_{d+1} \times_{\text{w}} {\CM}_{D}$. The line element in (\ref{DSansatz}) provides a geometric realization of discrete scale invariance when the functions $A, B$ and $\zeta_{i}$ are non-trivial periodic functions of $w$. The geometry in (\ref{DSansatz}) has a well defined conformal boundary when 
		$\displaystyle{\mathop{\lim}_{w\rightarrow \infty} A(w, \theta_{k})\text{, }  \mathop{\lim}_{w\rightarrow \infty} B(w, \theta_{k}) \text{ and }}$ $\displaystyle{ \mathop{\lim}_{w\rightarrow \infty} \zeta_{i}(w, \theta_{k})  }$ are bounded and regular functions of $\theta_{k}$.
		The dual field theory lives on the conformal boundary which is $d+1$ dimensional Minkowski space. We can also infer from the behavior of warp factor that there is mixing between scale transformations and internal symmetries in the dual field theory.
		
		The holographic $c-$theorem does not apply to such solutions due to the non-trivial dependence of warp factors on the coordinates of the internal manifold. There are few $AdS_{d+1} \times_{\text{w}} {\CM}_{D}$ solutions of 10$D$ and 11$D$ supergravity, for which there is a consistent Kaluza-Klein reduction on $\CM_{D}$ to $d+1$ dimensions ({\it e.g.}, \cite{Gauntlett:2007ma, Cvetic:2000yp} and references therein). However, not all solutions of the form (\ref{DSansatz}) admit a consistent truncation. A non-singular solution in higher dimension could appear singular in lower dimensions, even if a consistent truncation exists. So this does not violate the corollary of lower dimensional holographic $c-$theorem discussed earlier. 
		
		In the next section, we will find a solution of a 6D gravitational theory that is a warped product of $AdS_{5}$ and a circle. We will show that the solution is free of curvature singularities and regular. In section 3, we study the linearized fluctuations around this background and show that all modes are stable. In other words, there is no violation of ``Breitenlohner-Freedman type'' stability criterion. We provide arguments to show that the dual theory is not a soft deformation of a conformal field theory and also discuss a plausible UV completion. In section 4, we present solutions of type II supergravity that exhibit discrete scale invariance. These solutions are obtained using a solution generating technique similar to the TsT transformations discussed in \cite{Lunin:2005jy}. In section 5, we conclude with a discussion on the results of this paper.

		\section{A solution with discrete scale invariance}
	
		In this section, we present a solution of a 6D gravitational theory that has the right properties to be dual to a 3+1 dimensional discrete scale invariant theory. The 6D gravitational theory is described by the following action.  
		\be S_{6D} = {1\over 2 \kappa_{6}^{2}} \int d^6 x \( R - {1\over 2} \left(\del \chi \)^2 - V\(\chi\) \) \label{6Daction} \ee
		where $V\(\chi\) =\Lambda_{1}+\Lambda_2 \cos(\chi/\sqrt{2}) $, is the potential for the scalar field $\chi$. For convenience, we express $\Lambda_1$ and $\Lambda_2$ in terms of a characteristic length $L$ and a non-dimensional parameter $\alpha$ as follows:
		\be \Lambda_1 = -{8\over L^2} \( {2-\alpha^2 \over 1- \alpha^2}\), \quad \Lambda_2 = {12\over L^2} \( {\alpha^2 \over 1- \alpha^2}\) 
	\text{ with }0 \le \alpha < \sqrt{2/3}. \ee
	The reasons for choosing a specific range for $\alpha$ will become clear later.
	We will assume that $\chi$ is a compact scalar field ($\chi \equiv \chi + 2\pi \sqrt{2}$) but, in general, $\chi$ can have an unbounded field range. In string compactifications, the potential for angular moduli, such as axions, develops a periodic behavior from instanton corrections. The non-trivial periodic potential for the axion breaks the continuous axion shift symmetry to discrete shift symmetry. 
		
	%
	%
	
	 We are interested in finding a solution of this theory whose metric is a warped product of $AdS_{5}$ and a circle. The line element describing the geometry of such a solution takes the following form:
	 \be ds^2_{6D} = e^{2 C(w, \theta)} \left[ e^{2 w/L} \left( -dt^2 + d\vec{x}^2 \right)+ dw^2 \right] + e^{2 B(w, \theta)} \left( d\theta + {\CA}(w, \theta) dw\right)^2 \label{6Dmetric}\ee
	  A non-trivial axion flux is required to prevent the circle from shrinking to zero size. The warp factor must be periodic along the compact direction for the metric to be single valued and free of discontinuities. Also, the warp factor must have a non-trivial dependence on the compact direction when $\Lambda_2$ is non-zero. We would like to emphasize that null energy condition is not violated in this holographic description of discrete scale invariant theories. 
	
		 We will now present a discrete scale invariant solution of the gravitational theory in (\ref{6Daction}) with the above mentioned properties. The geometry of this solution is described by the line element in (\ref{6Dmetric}) with
		\be e^{2C(w,\theta)} = 1- \alpha^{2} \( \jsn(w/h, \alpha)\jcn\(N_a\theta, \alpha\) + \jcn(w/h, \alpha)\jsn(N_a\theta, \alpha)\)^2, \label{eqforC}\ee \be e^{2B(w, \theta)}= N_a^{2}L^2 \(1-\alpha^2\) \jdn^2(N_a \theta, \alpha)e^{-2 C(w,\theta)}, \quad {\cal A}(w,\theta) = {1\over N_ah}{\jdn(w/h, \alpha)\over  \jdn(N_a \theta, \alpha)}. \label{eqforB}\ee
	where $\jsn$, $\jcn$ and $\jdn$ are Jacobi elliptic functions (reparametrized to have periodicities $2\pi$, $2\pi$ and $\pi$ respectively). The definition and some properties of the Jacobi elliptic functions $\jsn$, $\jcn$ and $\jdn$ are listed in Appendix A. We will denote the metric described by (\ref{6Dmetric}), (\ref{eqforC}) and (\ref{eqforB}) by $g$. The axion equation of motion and Einstein's equations are solved by the following axion profile:
	\be \chi = 2\sqrt{2}\( \jam(N_a \theta, \alpha) + \jam(w/h, \alpha)\) \label{chi} \ee
	where $\jam$ is the Jacobi amplitude function (see Appendix A). Note that the partial derivatives of the axion field with respect to $w$ and $\theta$ are periodic. Hence, the contribution of the axion to the stress tensor is periodic.
	
	 It can be easily verified that the metric $g$ and the axion profile in equation (\ref{chi}) satisfy the equations of motion. This solution can be obtained by writing the equations of motion in first order form by choosing a suitable ansatz. In fact, there is a family of such solutions (with different potential for the scalar fields) that can be constructed by writing the equations of motion as first order equations. The precise details of these solutions and the first order formalism will appear in a future paper \cite{WIP}. 
	
	The parameter $h$ specifies the asymptotic behavior of the axion and $N_a$ determines the axion flux. When $N_a$ is zero, the circle degenerates and the geometry is not well-defined. However, it is possible to find asymptotically $AdS_{6}$ solutions with $N_a = 0$.

	
	  Let us now analyze the behavior of the metric $g$. Using the properties of the elliptic functions, it can be shown that $1 \ge e^{{2C(w, \theta)}} \ge 1- \alpha^{2}$, $ \(1- \alpha^{2}\)^{-1} \ge N_ah {\CA}(w, \theta) \ge  \(1- \alpha^{2}\)$  and $N_a^2 L^2 \ge e^{2B} \ge N_a^2 L^2 \(1- \alpha^{2}\)^2$.\footnote{See Appendix B for plots depicting the behavior of these functions.} This implies that the metric $g$
	 has a 3+1 dimensional conformal boundary when $w\rightarrow \infty$:
	\be ds^2_{bdy} = \mathop{\lim}_{w\rightarrow \infty} \Omega(w, \theta)^2 ds^2_{6D} = -dt^2 + d\vec{x}^2 \ee
	where $\Omega(w, \theta) = e^{-w/L} e^{-C}$. The dual field theory lives on the conformal boundary and it is invariant under 3+1-dimensional Lorentz transformations. Note that the size of the circle remains finite throughout the geometry indicating that the dual theory is gapless. The theory does not have continuous scale invariance but it is invariant under discrete scale transformations. Invariance under scale (discrete or continuous) transformations implies that there are arbitrarily low energy excitations or the theory is gapless. 
	Invariance of the metric in (\ref{6Dmetric}) under discrete scale transformation is realized as follows:{\footnote{When $\chi$ is non-compact, these transformations must be accompanied by a shift in $\theta$:
	$\displaystyle{\theta \rightarrow \theta + {2 n \pi \over N_a} }.$}}
	$$ t \rightarrow e^{2 n \pi h/L} t,\quad \vec{x} \rightarrow e^{2 n \pi h/L} {\vec{x}}, \quad w \rightarrow w - 2 n \pi h,\quad\text{for } n \in \mathbf{Z}.$$	
	Note that the invariance of the metric under discrete scale transformation is not an artifact of the choice of coordinates. This discrete scale invariance of the geometry can also be seen in coordinate independent quantities such as the Ricci scalar, which is given by
\be {\cal R} = -{10 (2+\alpha^2)\over L^2 (1- \alpha^2)} +{40\alpha^2 \over L^2 (1- \alpha^2)}  {\left[\jcn(N_a \theta, \alpha) \jsn(w/h, \alpha) +\jsn(N_a \theta, \alpha) \jcn(w/h, \alpha) \right]^2}\ee
	Note that the Ricci scalar, ${\cal R}$ is negative for all values of $w$ and $\theta$ when $0 \le \alpha < \sqrt{2/3}$. Scale invariance is broken if any curvature invariant depends on the holographic direction, $w$. When $h$ is finite, the Ricci scalar displays periodic behavior in $w$ and hence scale invariance is broken to discrete scale invariance.
	
	When $h \rightarrow \infty$, the axion becomes independent of $w$ and the solution can be written as
	 \be ds^2 = e^{2 \tilde{C}(\theta)} \left[ e^{2 w/L} \left( -dt^2 + d\vec{x}^2 \right)+ dw^2 \right] + e^{2 \tilde{B}(\theta)} d\theta^2, \quad \chi =  2\sqrt{2} \jam(N_a \theta, \alpha) \label{6Dmetric2}\ee
	 In this limit, all the curvature invariants are independent of $w$, and the geometry exhibits conformal invariance. In other words, the solution with $h\rightarrow \infty$ is dual to a conformal field theory. It is possible to find a choice of renormalization scheme in which this conformal field theory is described by a conformal RG limit cycle \LPR.\, Gravity duals of such conformal limit cycles can be obtained by performing a coordinate transformation on (\ref{6Dmetric2})  that just redefines the holographic direction $w$ in (\ref{6Dmetric2}): $w' = w  + {\cal W}(\theta, w)$ (where $\cal W$ is a periodic function of $w$ and $\theta$). This coordinate transformation results in a metric of the form (\ref{DSansatz}), but it neither modifies the behavior of the curvature invariants or the asymptotic behavior. Asymptotically $AdS$ solutions having the same boundary behavior (or same conformal boundary data) that are related by bulk diffeomorphisms are dual to the same theory (with different choices of the renormalization scheme). 

	Some diffeomorphisms in the bulk modify the conformal boundary data or the asymptotic behavior of certain fields in the gravity theory. Geometries related by such diffeomorphisms are dual to different theories. For instance, a coordinate transformations that modifies the periodicity of $\theta$ leads to a different solution with a different asymptotic behavior of the axion and the metric. This idea can be used as a solution generating trick for obtaining solutions satisfying different boundary conditions. In fact, it has been used in the past to generate black hole solutions from known black hole solutions (see for {\it e.g.,} \cite{Peet:2000hn, Hansen:2006wu}). Lunin and Maldacena \cite{Lunin:2005jy} showed that this technique combined with a sequence of T-dualities can be used to generate gravity duals that describe certain marginal deformations ($\beta$-deformation) of $\CN =4$ SYM theory. More recently, this transformation has been used to generate solutions of type II supergravity having non-relativistic symmetries \cite{NRCFT}. In Section 5 we will use this trick to generate solutions of 10D supergravity with discrete scale invariance. 

We will now show that $g$ is regular (non-singular) for all non-zero values of $N_a$ and $h$. First, we will show that all curvature invariants are finite.
	 The curvature invariants can be written as a polynomial (with appropriate contraction of indices) in $g^{\mu\nu}$ and derivatives of $g_{\mu \nu}$.  It follows from the discrete scale invariance of the metric that the Ricci scalar and other curvature invariants, such as $R_{ab}R^{ab}$,  $R_{abcd}R^{abcd}$, are all periodic functions in $w$ and $\theta$. These periodic functions can be expressed as a polynomial in $e^{2C}$, $e^{-2C}$, $e^{2B}$, $e^{-2B}$, ${\cal A}$ and their derivatives. All of these functions are regular and hence the curvature invariants must be regular periodic functions of $w$ and $\theta$. 
	
	We will now show that $g$ is geodesically complete. The evolution of null and timelike geodesics is described by the following equation
	\be e^{{2C}}\({dw\over d\lambda}\)^{2} + e^{-2C} e^{-2w/L}\( -E^{2} + \vec{p}^{2}\) + e^{2B} \({d\theta \over d\lambda} +  {{\cal A}}\({dw\over d\lambda}\)\)^{2} = k \ee
	where $k=0$ for null geodesics and $k=-1$ for timelike geodesics. 
In the above equation, $\lambda$ is the affine parameter and $(E, \vec{p})$ are the constants associated with the Killing vectors $(\del_{t}, \vec{\del})$. Using the fact that $e^{{2C}}$ is a bounded function, we can obtain the following inequality	 
$$ e^{{2C_{min} + 2C_{max}}}\({dw\over d\lambda}\)^{2} + e^{-2w/L}\( -E^{2} + \vec{p}^{2}\) \le -e^{2B + 2C_{max}} \({d\theta \over d\lambda} +  {{\cal A}}\({dw\over d\lambda}\)\)^{2} + ke^{2C_{max}} \le k $$
It is clear that neither null or timelike geodesics can reach the reach the conformal boundary $w = \infty$ in finite affine parameter. It appears that time-like and null geodesics can reach the Poincar\'e horizon at $w = -\infty$ in finite affine parameter. However, this is just the innocuous coordinate singularity associated with the Poincar\'e patch. This coordinate singularity can be removed by introducing global coordinates that cover the entire spacetime. For instance, we can simply use the usual transformation from $AdS$ Poncar\'e coordinates to global coordinates to obtain the following line element
$$ ds^2 = e^{2 {C}(\theta, w(\rho, T, \Omega))} \left[  \left( -\cosh^{2} \rho dT^2 + \sinh^{2} \rho d\Omega_{3}^{2} \right)+ d\rho^2 \right] + e^{2 {B}(\theta, w(\rho, T, \Omega))} (d\theta + {\cal{A}}(\theta, w(\rho, T, \Omega)))^2 $$
where $w(\rho, T, \Omega)$ relates the ``radial'' direction of the global coordinates to the radial direction in the Poincar\'e coordinates. We can again use the fact ${\cal A},~B,~C$ are bounded functions to show that the metric obtained after the transformation is geodesically complete. 
Hence, the solution that we have found in this section is regular for all non-zero values of $N_a$ and $1/h$.

\section{Linearized Fluctuations}
		
		In this section, we will analyze the linearized fluctuations around the background described by (\ref{6Dmetric}), (\ref{eqforC}), (\ref{eqforB}) and (\ref{chi}) and show that they are all stable. We will also use this analysis to argue that the dual field theory is not a soft deformation of a conformal field theory. In particular, we will show that the parameter $h$ is dangerously irrelevant which leads to a moderate hyperscaling violation that breaks continuous scaling symmetry to discrete scaling symmetry.{\footnote{ We would like to thank John McGreevy for numerous discussions on the interpretation of $h$ (boundary behavior of the axion) and about dangerous irrelevant operators. We would also like to thank Shamit Kachru for pointing out the connection with hyperscaling violation.}} 
		
		The linearized fluctuations of the metric can be decomposed into tensor, vector and scalar modes as follows:
		\be ds^2_{g'} = ds^2_g +h_{ab} dx^a dx^b + 2 h_{\theta a} dx^a \( d\theta + {\CA}(w,\theta)dw\)+ h_{\theta \theta}\( d\theta + {\CA}(w,\theta)dw\)^2 \ee
		 where the lower Latin indices denote the non-compact directions, $\gamma_{ab}$ is the metric on $AdS_5$ in Poincar\'e coordinates and 
		 \be h_{ab} =  \(h^{TT}_{ab} + \nabla_{(a} V_{b)} + S_T \gamma_{ab} + \(\nabla_a \nabla_b - {1\over 5} \gamma_{ab}\)S_L \).\ee
		  In the above expression, $\nabla$ denotes the covariant derivative in $AdS_5$. The fluctuations of the axion field ($\chif = \chi' - \chi$) couples to the scalar modes but not the vector or tensor modes. It is possible to eliminate the vector modes completely by choosing the longitudinal gauge where $h_{\theta a} =0$, $V_a =0$ and $S_L = 0$. In this gauge, only the traceless transverse tensor modes and the scalar modes needs to be considered. We will first analyze the traceless transverse tensor modes. 
		  
\bigskip

\noindent
	  {{{\it3.1. Stability of linearized fluctuations}}}	
	  
	  \bigskip
	  
		  The linearized Einstein equations for the traceless transverse modes decouple from the scalar modes and can be written as follows {\footnote{ In order to obtain the expression in (\ref{waveequation}) we have utilized the following fact:
		  
		 \centering{ $ \displaystyle{\( \del_{w} - {\cal A} \del_{\theta}\) \(\jsn(N_{a}\theta,\alpha)\jcn(w/h,\alpha) + \jsn(w/h,\alpha)\jcn(N_{a}\theta,\alpha)\) = 0} $
		  }}}
		\be \({e^{-5C-B}\over L^{2}\(1-\alpha^{2}\)}\del_\theta \( e^{5C-B}\del_{\theta} \)+ {\pounds} +e^{-2C} e^{-2 w/L} \square\right) H_{ab}=0, \label{waveequation} \ee
		where $H_{ab} = \Omega^2 h_{ab}^{TT}$, $\displaystyle\square$ denotes the d'Alembertian operator of 3+1 dimensional Minkowski space and ${\pounds}$ is a linear operator given by 
		\be  {\pounds} = e^{-2C}e^{-4w/L}\(\del_w - \cal{A} \del_{\theta}\)\left[e^{4w/L} \(\del_w - \cal{A} \del_{\theta}\)\right] \ee
The operator appearing in equation (\ref{waveequation}) is the Laplacian operator associated with the metric $g$. In other words, $H_{ab}$ satisfies the wave equation for a probe scalar in this background. 

Translation invariance along ($t,\vec{x}$) allows us to use Fourier decompoition along these directions: $\tilde{H}_{ab} = \int d^{3+1}x e^{i p.x} {H}_{ab}$. The linearized equation for $\tilde{H}_{ab}$ can be written in Schr\"odinger form by defining $\tilde{H}_{ab} = e^{-3C-2w/L}{\mathfrak{H}}_{ab}$. Then, equation (\ref{waveequation}) can be written in terms of $\mathfrak{H}_{ab}$ as follows
		 \be \(e^{-B}{\CQ}^{\dagger}_{1}{\CQ}_{1} + e^{-4C - B}{\CQ}^{\dagger}_{2}{\CQ}_{2} + {4\over L^{2}}e^{-3C}\)\mathfrak{H}_{ab} = e^{-3C} e^{-2 w/L} p^{2} \mathfrak{H}_{ab}, \label{SchrodingerTT}\ee
		 where $p^{2} = E^{2} - \vec{p}^{2}$ and the operators $\CQ_{1}$ and $\CQ_{2}$ are given by,
		\be \CQ_{1} = e^{-(B+C)/2} \left( -\del_{\theta} + 3 \del_{\theta} C \right), \quad
		 \CQ_{2} =  {e^{(B+C)/2}}\left( -\del_{w} +  {\cal A}\del_{\theta} \right)\ee
		In the above equations $\CQ^{\dagger}$ denotes the adjoint of $\CQ$ in the space of square integrable functions. If we assume a flat space norm for $\mathfrak{H}_{ab}$, then the left hand side of (\ref{SchrodingerTT}) is semi-positive definite. In other words, there are {\it no} square integrable solutions of (\ref{SchrodingerTT}) with negative values of $p^2$. This implies that the traceless-transverse tensor fluctuations are linearly stable. 
		
		We will now show that square integrability of $\mathfrak{H}_{ab}$ is equivalent to finiteness of the energy functional associated with the traceless-transverse modes. Finiteness of the energy functional associated with a field configuration defines a notion of {\it normalizablility} of the field configuration in curved spacetime. The normalizability condition for linearized fluctuations with stress tensor $T^{lin}_{AB}$ is given by
				\be {\cal E} = \int_{\Sigma} d^3\vec{x} dwd\theta \sqrt{\sigma}T^{lin}_{tA}\xi^A < \infty \ee
		where $\Sigma$ denotes a constant time slice with induced metric $\sigma$, $\xi^A$ is a time-like Killing vector and $A\in \{t,\vec{x}, w, \theta\}$. Note that we have assumed that there is no energy-momentum flux leaking out of the boundary at $w= \infty$. In order to compute $T^{lin}_{AB}$, we can treat $H_{ab}$ as a probe massless scalar. This is justified as $H_{ab}$ is governed by a massless scale wave equation. Without loss of generality, we can assume $\xi = \del_t$; in this case, it is sufficient to compute $T^{lin}_{tt}$ component of the energy momentum tensor. Now, the finiteness condition on energy simplifies to 
		\be \int_{\Sigma} d^3\vec{x} dwd\theta e^{3C}e^{2w/L} \left(\del_t {H}_{ab} \right)\left(\del_t {H}_{cd}\right)\eta^{ac}\eta^{bd} < \infty \implies \int d^3 p dwd\theta \mathfrak{H}_{ab}\mathfrak{H}_{cd}\eta^{ac}\eta^{bd} < \infty \ee
		This shows that the finiteness condition on energy of the linearized fluctuations is equivalent to square integrability condition on $\mathfrak{H}_{ab}$. This completes our discussion on linear stability of transverse traceless modes. 
		
	  We will now show that the linearized fluctuations of the scalars are also stable. The axion fluctuation $\chif$ and the radion fluctuation $h_\theta^\theta$ are determined through constraint equations if $S_T$ is known. Using linearized Einstien's equations and axion equation of motion we can show that $S_T$ satisfies the following equation:
	  \be \({e^{C-B}\over L^{2}\(1-\alpha^{2}\)}\del_\theta \( e^{C-B}\del_{\theta} \)+ e^{2C}{\pounds} + e^{-2 w/L} \square - U\(w,\theta\)\right) S_{T}=0, \label{scalarequation} \ee
	  where, 
	  \be U\(w,\theta\) =  \({12\over L^2 \(1-\alpha^2\)}-{(2 - \alpha^2)\over L^2 \(1-\alpha^2\)} \(8  - 6 e^{2C(w,\theta)} \) \)\ee
	  Note that $U\(w,\theta\)$ is positive for all $\alpha < 1$. We can write (\ref{scalarequation}) in the Schr\"odinger form by changing the independent variable to $\mathcal{S}_T = e^{C}e^{2w/L} S_T$. The linearized equation for scalar fluctuations in the Schr\"dingier form can be written as
		 \be \(e^{3C-B}{\CQ}^{\dagger}_{3}{\CQ}_{3} + e^{-B-C}{\CQ}^{\dagger}_{2}{\CQ}_{2} +{4\over L^{2}}+ U\(w, \theta\)\)\mathcal{S}_{T} =  e^{-2 w/L} p^{2} \mathcal{S}_{T}, \label{SchrodingerScalar}\ee
We can repeat the earlier arguments to show that the left hand side of the above equation is positive definite. Hence, all finite energy configurations of $\mathcal{S}_T$ must have positive $p^2$. In other words, the scalar modes are also linearly stable. This implies that all linearized fluctuations around $g$ are stable.

	  \bigskip 

\noindent
	  {{{\it3.2. Dangerously irrelevant operators}}}	
	  
	  \bigskip
	  We will now argue that the dual field theory is not a soft deformation of a 3+1 dimensional CFT. We know that when $h\rightarrow \infty$ the dual field theory is a conformal field theory. For large, but finite values of $h$ we can assume that the UV description of the dual theory is a deformation of a conformal field theory by $\sum_{\Delta}\lambda_{\Delta}(h) {\CO}_{\Delta}$, where $\lambda_\Delta$ denote couplings and the operators ${\cal O}_{\Delta}$ denote operators of the conformal field theory with dimension $\Delta$. In the limit when $h$ is large, these operators are dual to linearized fluctuations around the $h \rightarrow \infty$ solution in (\ref{6Dmetric2}). Let us analyze the nature of the operator dual to the scalar mode $S_T$. Using the $h \rightarrow \infty$ limit of (\ref{scalarequation}) or (\ref{SchrodingerScalar}), we can show that the dimension of the scalar operator satisfies the following eigenvalue equation
	  	\be   \(e^{3C-B}{\CQ}^{\dagger}_{3}{\CQ}_{3}  + U\(w, \theta\)\)\mathcal{S}_{T} = \Delta(\Delta - 4) {\cal S}_T \label{Dimension}\ee
We can use the positivity of the left hand side to conclude that $\Delta >4$. There is no relevant or marginal scalar deformation and hence the parameter $h$ corresponds to an irrelevant deformation. Hence, the dual field theory is not a soft deformation of a conformal field theory. The parameter $h$ leads to violation of continuous scaling laws and affects the IR physics even though it is irrelevant. In particular, the period of the limit cycle is determined by $h$. Hence, the parameter $h$ should be related to a dangerously irrelevant coupling. It is well known that dangerously irrelevant operators lead to hyperscaling violation in condensed matter systems (for \eg\, see \cite{Subir}).  The dual field theory can be considered as a special example of theories exhibiting hypercsaling violation where scale invariance is restored periodically resulting in discrete scale invariance. 

The dual DSI field theory (for any $h$) cannot be UV completed by a 3+1 dimensional CFT that is dual to an asymptotically $AdS_5$ solution of (\ref{6Daction}). This can be explained as follows. To describe the UV completion in the gravity theory, we have to find a saddle point of (\ref{6Daction}) that interpolates between a metric of the form (\ref{6Dmetric2}) as $w\rightarrow \infty$ (UV of the dual theory) and the metric $g$ as $w$ approaches $-\infty$ (IR of the dual theory). Note that the UV asymptotic ($w= \infty$) should satisfy the equations of motions as well. The operator dimensions of scalar deformations near $w = \infty$ is determined from (\ref{Dimension}). It is clear that the scalar deformations cannot be marginal or relevant at the UV. This suggests that the discrete scale invariant theory dual to $g$ cannot be UV completed by a 3+1 dimensional CFT with a classical gravity dual. The 6D action in (\ref{6Daction}) admits two inequivalent $AdS$ solutions when $\chi$ is a compact field. So, it might be possible to UV complete this theory by a theory on $\mathbf{R}^{3,1}\times_{w}\mathbf{S}^{1}$.  To show this, we have to find a solution interpolating between $AdS_{6}$ solution of (\ref{6Daction}) and $g$.

We will not find such a solution in this paper, but it would be interesting to find one in the future. We will argue that discrete scale invariant theories can be UV completed by finding alternate examples. This is done in the next section where we will present solutions of type II supergravity that are dual to discrete scale invariant theories. These theories can be UV completed (in principle) by an open string theory describing the worldvolume theory of certain $D$-branes. 

Before we proceed to discuss examples of discrete scale invariant solutions in type II supergravity, we will comment about the solutions of the wave equation in (\ref{waveequation}) and the behavior of correlation functions.

\bigskip

\noindent
	  {{{\it3.3. Comments on correlation functions and Bloch decomposition}}}	
	  
	  \bigskip	
	  \noindent	
		 The most general solution of (\ref{waveequation}) takes the following form{\footnote{To obtain the result in (\ref{SolLin}), we have utilized the result in footnote 5. We have also used some properties of Jacobi elliptic functions to express terms involving $e^{B}$ and derivatives of $C(w,\theta)$ as a function of $C(w, \theta)$.}}
		\be H_{ab} = \varepsilon^{TT}_{ab}\mathop{\sum}_{\Delta_B} {\Psi_{\Delta_B}}\(w,p^{2}\){\cal Y}_{\Delta_B}\(e^{2C(\theta, w)}\) \label{SolLin}\ee
		where $ \Psi_{\Delta_B}\(w,p^{2}\)$ is the solution of a massive scalar wave equation in $AdS_{5}$ with $m^{2} = \Delta_B(\Delta_B-4)$, $\varepsilon^{TT}_{ab}$ denotes transverse traceless polarization tensor and ${\cal Y}_{\Delta_{B}}(x)$ satisfies a Sturm-Liouville equation. Technically, $\Delta_B$ is not a good quantum number in discrete scale invariant theories. However, $\Delta_{B}$ can be associated with discrete scale transformations in the same sense Bloch momentum can be associated with discrete translation (spatial) invariance. 			
That is, the decomposition in (\ref{SolLin}) can be considered as a Bloch decomposition associated with discrete scale invariance. 

Note that most monotonicity theorems on RG flows based on field theoretic techniques assume that the UV theory can be expressed as a deformation of a CFT by operators (relevant) with ``well-defined'' conformal dimensions. This assumption may not apply to discrete scale invariant theories where $\Delta_B$ is not a good quantum number in the strict sense. This does not imply that operators cannot be classified as marginal, relevant or irrelevant. Wilson and G{\l}azek \cite{Wilson} demonstrated using a simple quantum mechanics example that it is possible to classify operators of a discrete scale invariant theory as marginal, relevant or irrelevant based on Wegner's eigenvalues \cite{Wegner}. 
At present, it is not clear how to extend the analysis in \cite{Wegner, Wilson} to the present context. However, it seems reasonable to assume that $\Delta_B$ (Bloch conformal dimension) can be used to classify operators (of the dual DSI theory) as marginal, relevant or irrelevant following the usual holographic dictionary. 

We will now analyze the behavior of correlators of a massless scalar field which can be considered as a proxy for $H_{xy}$.{\footnote{We will defer the actual computation for future work. We would like to thank Nikolay Bobev, Chris Herzog, John McGreevy and Balt van Rees for numerous discussions on this calculation and for explaining the difficulties involved.}}
The two-point function of scalars can be computed if the bulk-bulk propagator is known. The bulk-bulk propagator for a scalars is determined through the following equation
\be \({e^{-5C-B}\over L^{2}\(1-\alpha^{2}\)}\del_\theta \( e^{5C-B}\del_{\theta} \)+ {\pounds} +{e^{-2C}\over e^{2 w/L}} \square\right)G(w,w';x,x') ={1\over \sqrt{g}} \delta(w-w'){\delta}(x-x') \label{Green}\ee
 In principle, we can solve equation (\ref{Green}) to obtain the two point function of scalars.{\footnote{This can be done using numerical techniques such as boundary element method if it cannot be solved analytically.}} The bulk-boundary propagator can be determined from $G(w,w';x,x')$ using Green's second identity which gives the two-point functions. 
 
  We can verify that the bulk-bulk propagator cannot be expressed as a function of $AdS$ geodesic distance; in particular, it is not invariant under continuous scale transformations. This implies that the bulk-boundary propagator, does not respect conformal symmetries or continuous scaling symmetry. However, $G(w,w';x,x')$ is invariant under $(w,w';x,x') \rightarrow (w+2 n\pi h, w'+2 n\pi h;e^{2 n\pi h/L}x,e^{2 n\pi h/L}x')$. This implies that the bulk-boundary propagator should exhibit log-periodic behavior in $|x-x'|$ or in momentum scale. 
We have also shown that the linearized fluctuations are stable and there is no violation of Breitlohner-Freedman (BF) stability criterion. Hence, there is no indication that the dual field theory is non-unitary. The log-periodic behavior exhibited by correlation functions is associated with the Bloch decomposition described earlier. 
 Note that log-periodic behavior is not associated with complex values of $\Delta_B$ unlike the operators in non-unitary CFTs dual to BF violating mass in $AdS$. In such non-unitary CFTs, the log-periodic behavior respects conformal invariance and can be determined using the conformal algebra.

\section{Solutions of type II supergravity}

In this section, we will make use of a solution generating technique to find solutions of type II supergravity exhibiting discrete scale invariance. This provides some hope of finding conventional field theoretic descriptions of theories exhibiting discrete scale invariance. The solution generating technique used here is a simple generalization of Lunin-Maldacena TsT transformation \cite{Lunin:2005jy}. This technique allows us to generate new solutions of type IIB supergravity from an already known solution of type IIB supergravity. In this paper, we will generate DSI solutions starting with $AdS_5 \times \mathbf{S}^5$ solution of type IIB supergravity. The metric on $AdS_5 \times \mathbf{S}^5$ can be written as follows:
\be ds^2  = e^{2w/L}\(-dt^2 + d\vec{x}^2\) + dw^2 + L^2 ds^2_{\mathbf{S}^5} \ee
We write $\mathbf{S}^5$ as a Hopf vibration over $\mathbf{CP}^2$ as follows (see Appendix C for details):
\be ds^2_{\mathbf{S}^5} = d\varphi^2 + d\mu^2 +  \sin^2\mu\left[\( d\psi + \cos \theta d\phi  \)  d\varphi +{1\over 4} \(d\theta^2 + d\phi^2 + d\psi^2 + 2 \cos\theta d\psi d\phi \)\right] \ee
The solution generating trick involves the following three sequential steps: (1) T-duality along $\varphi$ (Hopf circle), (2) coordinate transformation: $\phi \rightarrow \phi + \beta \varphi + \alpha h f(w/h)$ and (3) T-duality along $\varphi$. Note that the second step modifies the conformal boundary data; in particular it modifies the periodicity of $\phi$ and resulting in a shift in the conserved charge associated with $\varphi$ translation symmetry. In addition,this shift is scale dependent when $\alpha$ is non-trivial. Note that this set of operations produces a new solution of the equations of motion from an already known solution in the same way, but with different asymptotic boundary data. We will assume that $f'(x)$ is a bounded periodic function (with period $2\pi$) and we will denote it by $F_p(x)$. Under the action of the above set of transformations, the line element (string frame) is modified as follows:
\be ds^2_{\text{string}}  =  e^{2w/L}\(-dt^2 + d\vec{x}^2\) + e^{2\gamma(\theta,\mu,w)}dw^2 + L^2 ds^2_{\mathcal{M}^5} + 2 \zeta_\varphi dw d\varphi + 2 \zeta_\phi dw d\phi + 2 \zeta_\psi dw d\psi  \ee
where,
$$ e^{2\gamma(\theta,\mu,z)} = 1 + {\alpha^2 L^2 \left[F_p\({w/ h}\)\right]^{2} \sin^2 \mu \over 4 \kappa}, \quad \kappa = \(1+ {\beta^2 L^4\ \sin^{2}\mu \over 4}\( \sin^{2}\theta + \cos^{2}\mu \cos^{2}\theta\) \)$$
$$ ds^2_{\mathcal{M}^5} = {1\over \kappa} d\varphi^2 + d\mu^2 + { \sin^2\mu\over \kappa}\left[\( d\psi + \cos \theta d\phi  \)  d\varphi +{1\over 4} \(\kappa d\theta^2 + d\phi^2 + d\psi^2 + 2 \cos\theta d\psi d\phi \)\right] $$
$$ \zeta_{\varphi} = {\alpha L^{2} \over 2 \kappa} \left[\(F_p\({w/h}\)\right] \sin^{2}\mu\cos \theta\), \quad  \zeta_{\psi} = {\alpha L^{2} \over 4 \kappa} \(\left[F_p\({w/ h}\)\right] \sin^{2}\mu\),\quad \zeta_{\phi} = {1\over 2}\zeta_{\varphi}$$
Note that $g_{ww}$, $\zeta_{\varphi}$, $\zeta_{\phi}$ and $\zeta_{\psi}$ are invariant under only discrete shifts in $w$. In other words, the metric exhibits discrete scale invariance. We can aslo show that this is not a coordinate artifact  by evalutaing curvature invariants. The dilaton ($\Phi$) and the NS 2-form field ($B$) of type IIB supergravity have non-trivial profiles in this solution.  The dilaton profile is given by $e^{-2 \Phi} = e^{-2 \Phi_{0}}\kappa$. The non-trivial components of the NS 2-form field are:
 $$ B_{\varphi \psi} ={ \beta L^{4} \sin^{2}\mu \cos^{2}\mu \cos \theta \over 4 \kappa}, \quad B_{\varphi \phi} ={ \beta L^{4} [F_{p}\(w/h\)]\sin^{2}\mu\(\sin^{2}\theta+ \cos^{2}\mu \cos^{2} \theta\) \over 4 \kappa} $$
 $$ B_{\phi \psi} = {\beta L^{4} \sin^{4}\mu \sin^{2} \theta\over 8 \kappa}, \quad B_{\varphi w} = \alpha L^{2} [F_{p}\(w/h\)]B_{\varphi \phi}, \quad B_{w \psi} = \alpha L^{2} [F_{p}\(w/h\)]B_{\phi \psi} $$
Note that the Einstein frame metric can be obtained from the string frame metric by multiplying by $e^{-\Phi/2}$. 

When $\alpha =0$, the dual field theory is a deformation of ${\cal N}=4$ SYM theory arising from replacing products of fields in the Lagrangian by star products \cite{Ganor, Lunin:2005jy}. However, the star product only introduces certain phases in the terms present in the Lagrangian. This phase is related to the deformation parameter $\beta$. When $\alpha \neq 0$, in addition to this phase becoming scale dependent, the conformal dimensions of the operators are also shifted by a function of the $R$-charges. At present, it seems difficult to translate these operations into field theoretic language and leave this for future. 
In this regard, it seems worthwhile to construct solutions that preserve supersymmetry by performing the generalized TsT transformation on a different set of $U(1)$ directions.\footnote{We would like to thank Nikolay Bobev for this comment.} In addition, it might be possible to use non-renormalization theorems to address questions about $1/N$ corrections. 

Note that the above solution generating trick can be used to generate a large class of solutions breaking various space-time or internal symmetries. In particular, it is straightforward to construct non-relativistic examples of discrete scale invariant theories by choosing a null direction instead of $\phi$ in the generalized TsT transformation \cite{NRCFT}. This procedure might be helpful in finding black hole solutions that describe discrete scale invariant theories at finite temperature. It would be interesting to know how various transport properties behave in discrete scale invariant theories. 

\section{Comments and Discussions}

In this note, we have used the holographic principle to argue for the existence of Lorentz invariant, discrete scale invariant theories that are described by limit cycle RG flows. However, this does not imply the absence of a function that is monotonic along the RG flow. In other words, there is a sense in which the number of degrees of freedom decreases as the theory flows towards the IR.{\footnote{We would like to thank Suresh Govindarajan, Sean Hartnoll and Diego Hofman for discussions on this.}} This fact can be easily seen on the gravity side using Raychaudhuri equation. The Raychaudhuri equation translates null energy condition into a monotonicity condition on the expansion parameter for null congruences. In the case of RG flows between CFTs, it can be shown that the monotonicity of the expansion parameter implies the $a-$function is monotonic \cite{Alvarez:1998wr, Sahakian:1999bd}. Such an $a-$function is not defined for discrete scale invariant theories and it is not clear how to construct a monotonic function of couplings using the information about the monotonicity of the expansion parameter. However, this indicates that it might be possible to set an arrow on RG flows even if it exhibits cyclic behavior.

The existence of gravity duals for cyclic RG flows suggests that there is a landscape of discrete scale invariant solutions of type IIB supergravity that cannot be described using a conventional lower dimensional effective action of the form (\ref{DWaction}). It would be interesting to know if a lower dimensional framework exists for studying such solutions. 

This also suggests the possible existence of tunneling transitions that are not described by conventional Coleman-de Luccia instantons. In particular, the lower dimensional effective theory framework used to describe Coleman deLuccia instantons cannot describe tunneling transitions that involve mixing of an arbitrarily large number of Kaluza-Klein modes \cite{WIP}. {\footnote{ In fact, the present work on cyclic RG flows started with questions related to false vacuum decay into Anti-de Sitter space.}}

The existence of gravity duals for cyclic RG flows also raises the following question: Is it possible to find classically stable and geodesically complete, bouncing cosmological solutions of general relativity without relaxing the conditions on energy-momentum tensor?{\footnote{ Graham \etal \cite{Graham:2011nb} showed that such solutions exist if the conditions on the energy-momentum tensor is relaxed.}} In particular, can we find a solution of the following form, exhibiting periodic behavior in time without relaxing the energy conditions \cite{WIPSujan}
\be ds^2 = e^{2 A(t, \theta_{k})} \left[  \left( -dt^2 + d\Sigma^2 \right) \right] + e^{2B(t,\theta_{k})}g_{i j} d\theta^{i} d\theta^{j} + 2 \zeta_{i}(t,\theta_{k}) dt d\theta^{i} \label{Cosmology} \ee
 This possibility seems to be ruled out by Hawking-Penrose singularity theorems \cite{Hawking}. However, the assumptions of Hawking-Penrose theorems need not apply to solutions of the above form. In particular, it is not clear if the above solution contains a closed trapped surface, which is assumed to exist for proving Hawking-Penrose singularity theorems. Another concern that needs to be addressed is the following: How can such solutions be consistent with second law of thermodynamics?{\footnote{We would like to thank Leonard Susskind for raising this question.}} At present, the answer to this question is not clear, but this question is tied to the microscopic details of entropy production and how entropy is actually defined.{\footnote{We would like to thank Peter Graham, Shamit Kachru and Surjeet Rajendran for detailed discussions on this topic.}} Also, there is no evidence that observers in such a universe would experience the second law as correct; in particular, it's not clear why there should be any low-entropy initial state.{\footnote{ We would like to thank John McGreevy for suggesting this.}} However, it seems that if the strong energy condition is satisfied, there exists a notion of time's arrow that can be defined using Raychaudhuri equation. It is not clear if the thermodynamic arrow of time is actually related to this. It is also not clear, how quantum effects modify the singularity theorems. So a classical bouncing solution that is geodesically complete and stable could be unstable quantum mechanically \cite{Graham:2011nb}.

		\section*{Acknowledgements}
		It is a great pleasure to thank A. Adams, D. Anninos, N. Bobev, S. Dabholkar, E. Dyer, X. Dong, P. Gao, S. Govindarajan, P. Graham, D. Harlow, S. Hartnoll, C. Herzog, D. Hofman, S. Kachru, J. H. Lee, H. Liu, R. Loganayagam, R. Mahajan, J. McGreevy, M. Mezei, D. S. Park, J. Polchinski, S. Pufu, S. Rajendran, L. Rastelli, M. Roberts, M. Rocek, E. Shaghoulian, S. Shenker, E. Silverstein L. Susskind, J. Thaler and B. van Rees for valuable comments and discussions. We would like to thank N. Bobev, C. Herzog, R. Loganayagam, J. McGreevy and J. Polchinski for their valuable comments on the draft.  We are grateful to Shamit Kachru for helpful discussions on bouncing cosmological solutions and also for his comments on the present work. We would especially like to thank Chris Herzog and John McGreevy for their comments on earlier ideas that were not well-formualted and for their support and encouragement. This work was partially supported by the funds provided by National Science Foundation under NSF Grant No. PHY-0969739 and NSF Grant  No. PHY-0844827. Any opinions,
findings, and conclusions or recommendations expressed in this material are those of the
authors and do not necessarily reflect the views of the National Science Foundation.
		\appendix
		\section{Jacobi Elliptic functions}		
		In this appendix, we list the properties of the reparametrized functions $\jam,~\jdn,~\jcn$ and $\jsn$. The reparametrized Jacobi elliptic functions are defined as follows:
		\be \jam(x,\alpha) = \text{am}(2K x/\pi,\alpha), \quad  \jdn(x,\alpha) = \text{dn}(2K x/\pi,\alpha), \ee 
		\be \quad \jsn(x,\alpha) = \text{sn}(2K x/\pi,\alpha), \quad  \jcn(x,\alpha) = \text{cn}(2K x/\pi,\alpha)\ee
		where, $K$ is the complete elliptic integral of the first kind:
		\be K = {\mathcal{K}}(\alpha) = \int_0^{\pi/2} {dx' \over 1- \alpha^2 \sin^2(x')}. \ee
		The function $\text{am}(z,\alpha)$ satisfies the following differential equation:
		\be\( {dy\over dz}\)^2 = 1- \alpha^2 \sin^2 y\ee
		The remaining elliptic functions are defined as follows:
		\be \text{dn}(z,\alpha) = {d\over dz} \text{am}(z, \alpha)\ee
		\be \text{dn}^2(z, \alpha) = 1-\alpha^2 \text{sn}^2(z, \alpha)  \ee
		\be \text{cn}^2(z, \alpha) + \text{sn}^2(z, \alpha) = 1\ee
		The elliptic functions can be written as a Fourier series as follows
		\be \text{dn}(z,\alpha) ={\pi\over 2K }+{2\pi\over K \alpha} \mathop{\sum}_{n=1}^\infty {q^n\over 1+q^{2n}} \cos\( {2n\pi z\over 2K}\)  \ee
		\be \text{cn}(z,\alpha) ={2\pi\over K \alpha} \mathop{\sum}_{n=0}^\infty {q^{n+1/2}\over 1+q^{2n+1}} \cos\( {(2n + 1)\pi z\over 2K}\)  \ee
		\be \text{sn}(z,\alpha) ={2\pi\over K \alpha} \mathop{\sum}_{n=0}^\infty {q^{n+1/2}\over 1 - q^{2n+1}} \sin\( {(2n + 1)\pi z\over 2K}\)  \ee
		where $q = \exp(-\pi \CK(1-\alpha)/K)$ is the nome. Note that the functions $\jdn$, $\jsn$ and $\jcn$ have periods $\pi$, $2 \pi$ and $2\pi$. Further, it can be shown that the function $\text{dn}$ is bounded from below by $(1-\alpha^2)^{1/2}$ and bounded above by $1$. While $\text{cn}$ and $\text{sn}$ are bounded above and below by $1$ and $-1$. 
		\section{ Plots showing the behavior of $e^{2C}$, $e^{2B}$, ${\cal A}$ \& $\chi$ }
		This appendix shows the behavior of the functions $e^{2C}$, $e^{2B}$, ${\cal A}$ \& $\chi$ when $\alpha^2 = 0.3$. The plot on the left hand side shows the behavior of these functions for fixed values of $\theta$ and the plot on the righthand side shows the behavior for fixed values of $z = w/h$.
		\vspace{.50cm}
		\begin{figure}[htbp]
		   \centering
		      \includegraphics[scale=0.45]{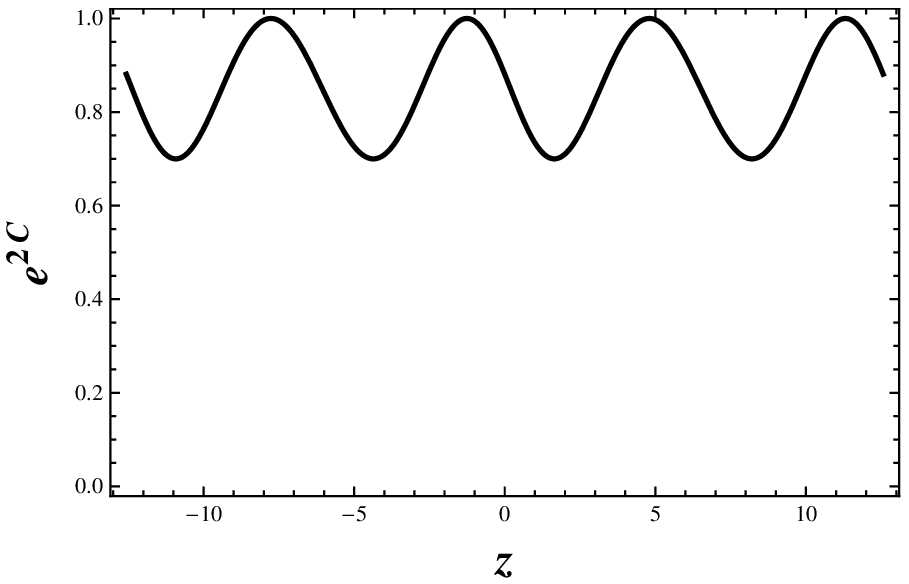} 
		       \hspace{.1cm}
		         \includegraphics[scale=0.35]{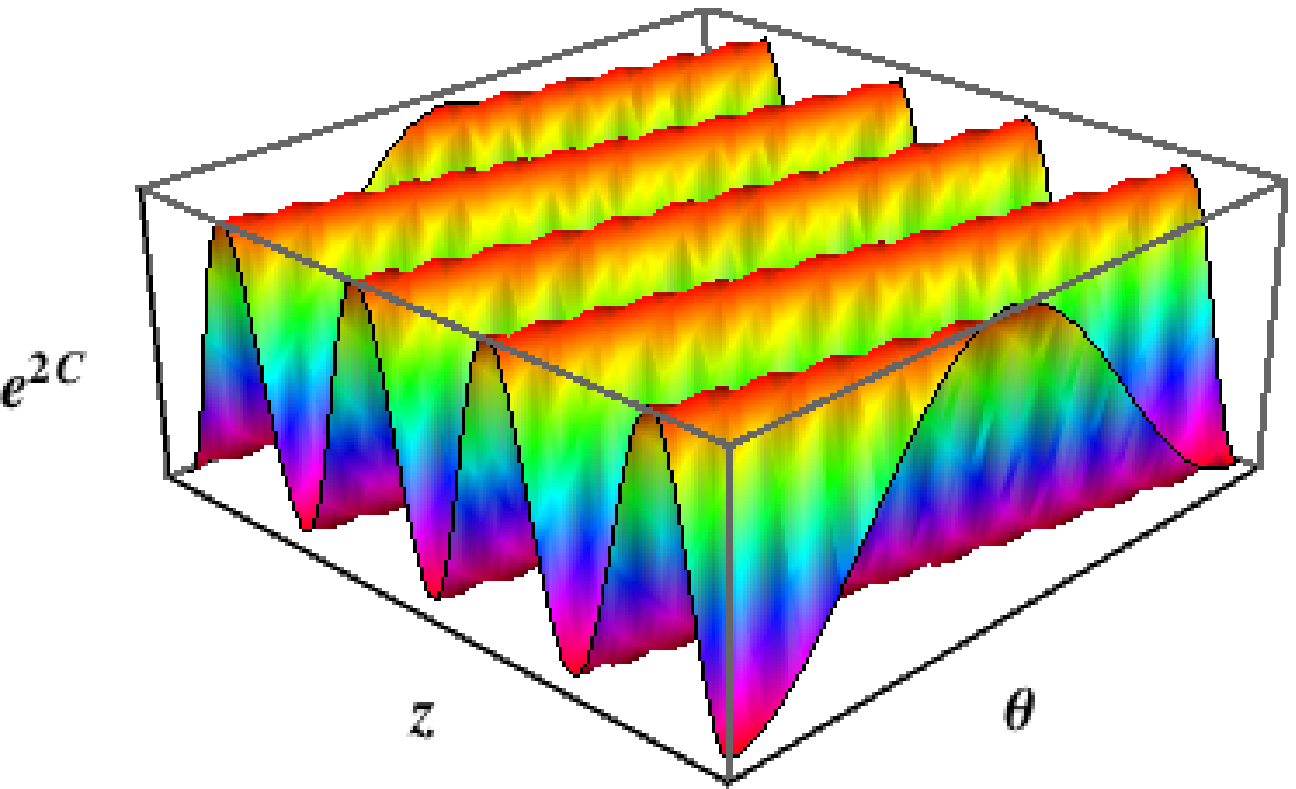}
		          \hspace{.1cm}
		      \includegraphics[scale=0.45]{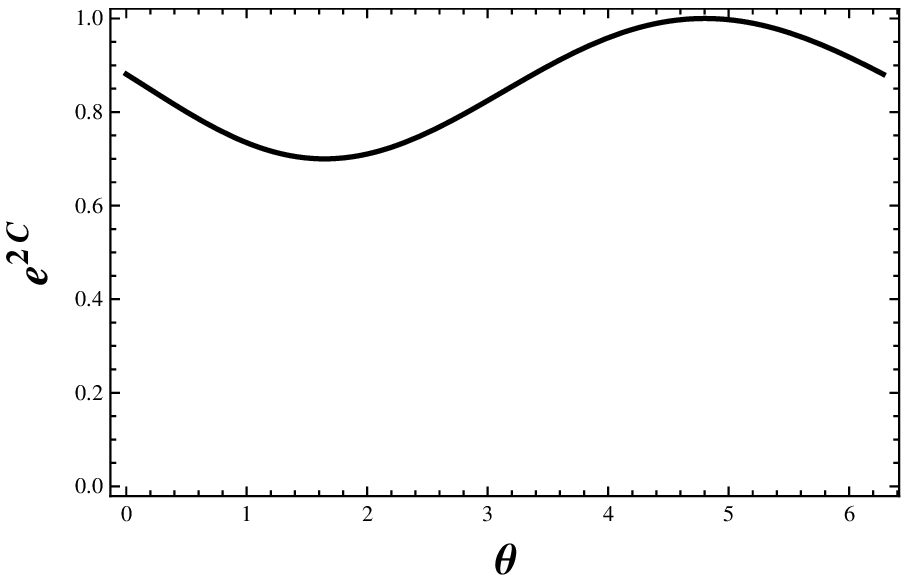} 
		   \caption{$e^{2C}$}
		   \label{fig:e2C3D}
		   \vspace{.5cm}
		   \end{figure}
		   \begin{figure}[htbp]
		   \centering
		      \includegraphics[scale=0.45]{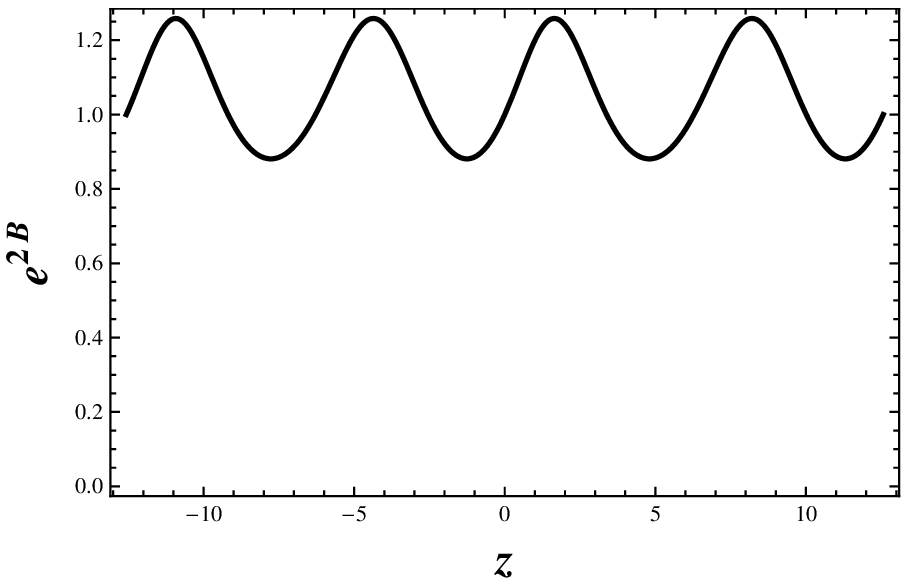} 
		       \hspace{.1cm}
		         \includegraphics[scale=0.35]{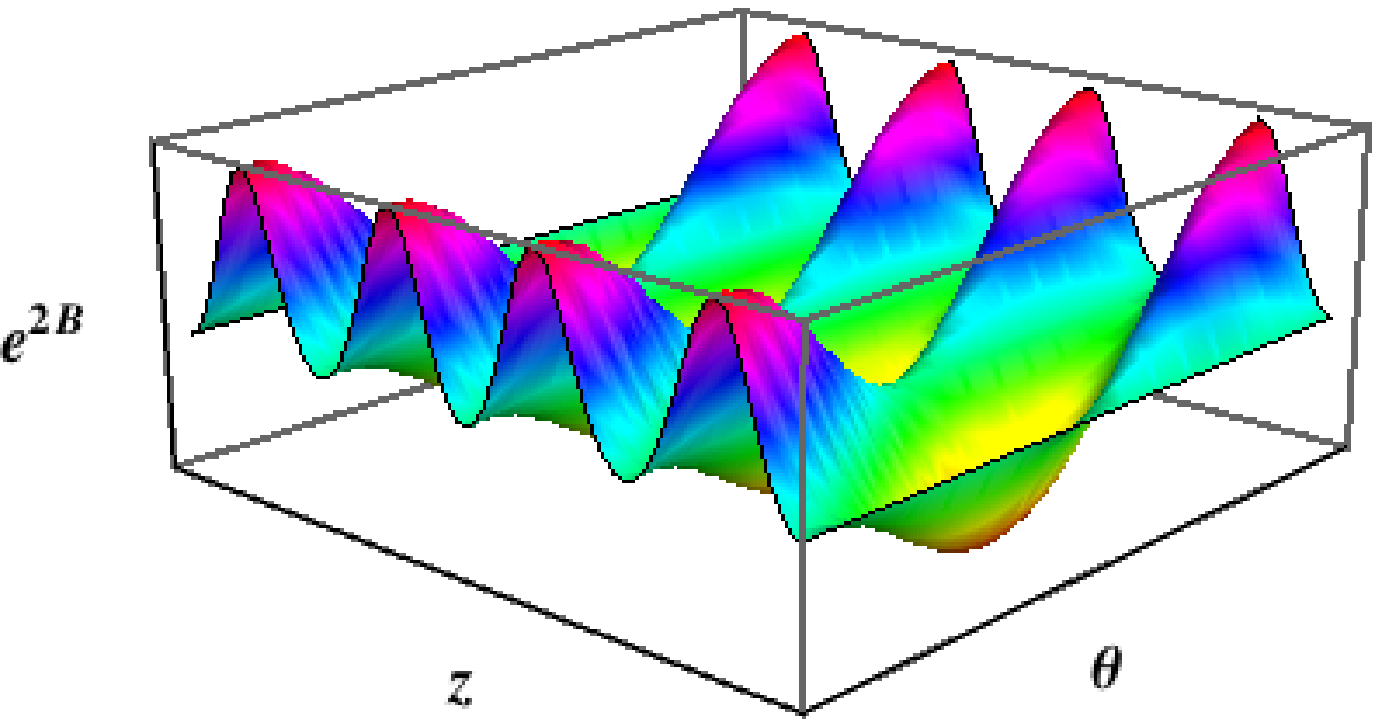} 
		          \hspace{.1cm}
		      \includegraphics[scale=0.45]{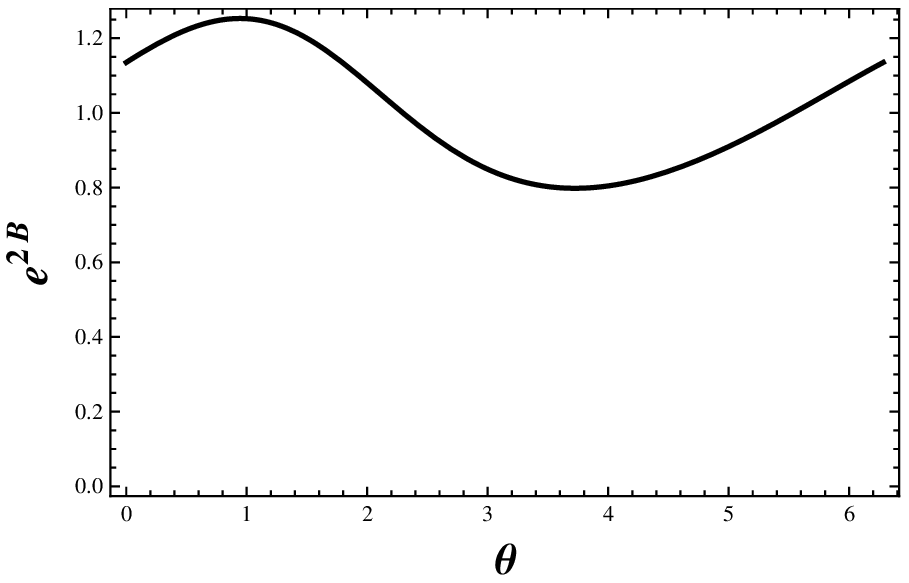} 
		   \caption{$e^{2B}$}
		   \label{fig:e2B3D}
		    \vspace{.50cm}	
		   \centering
		      \includegraphics[scale=0.45]{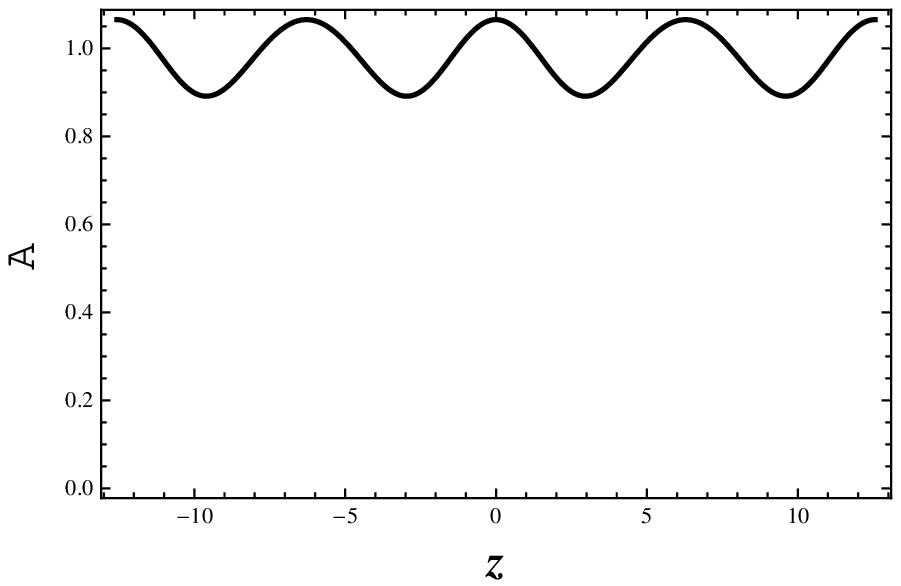} 
		       \hspace{.1cm}
		       \includegraphics[scale=0.35]{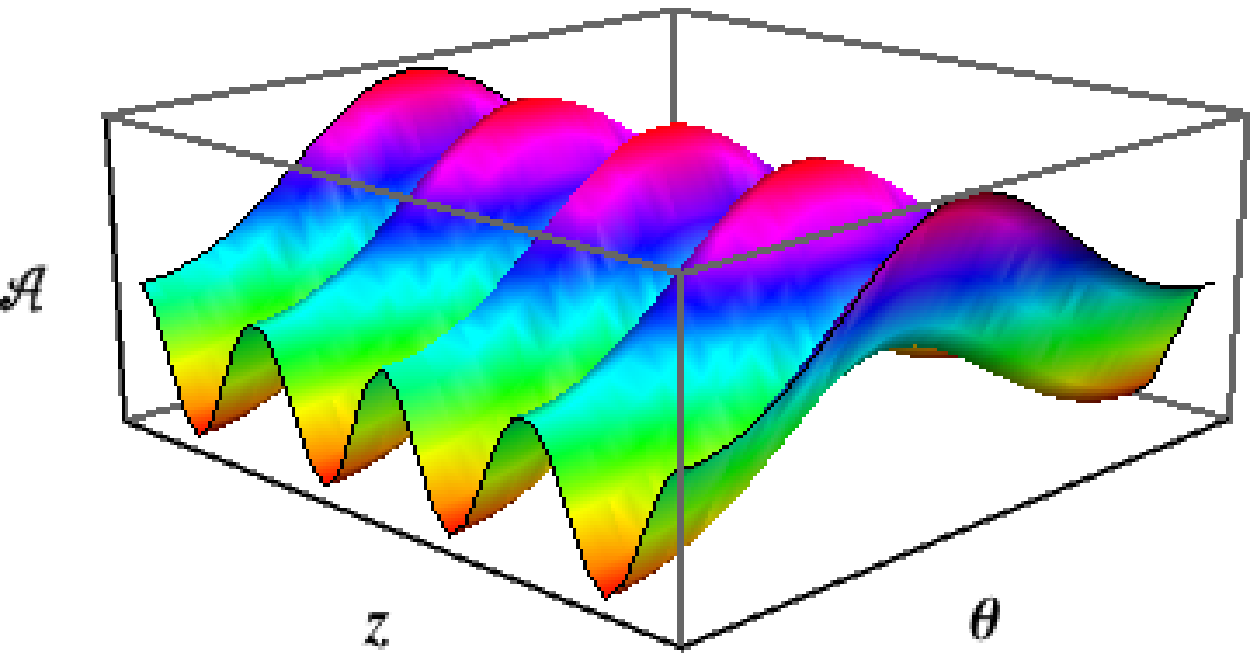} 
		        \hspace{.1cm}
		      \includegraphics[scale=0.45]{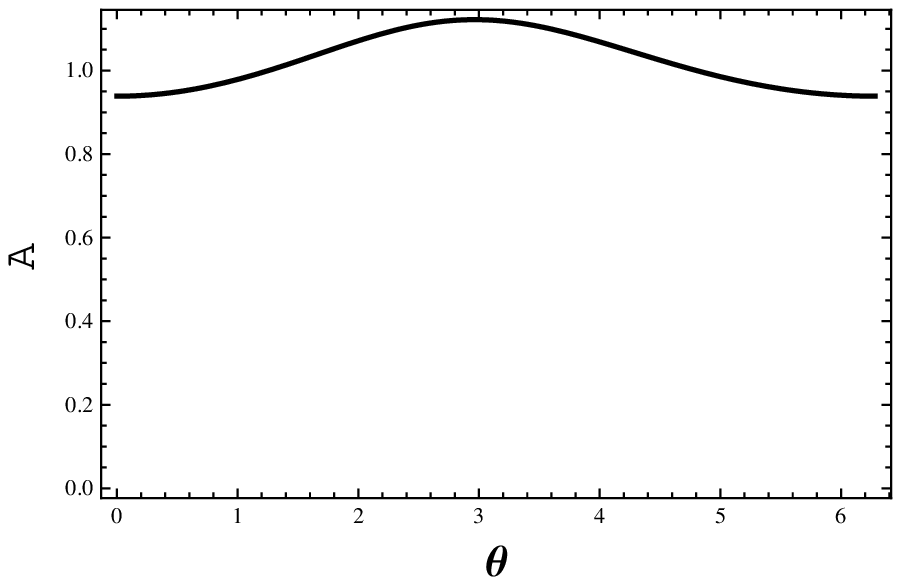} 
		   \caption{${\cal A}$}
		   \label{fig:AA3D}
		    \vspace{.5cm}
		     \end{figure}   
		   \begin{figure}[htbp]
		   \centering
		      \includegraphics[scale=0.45]{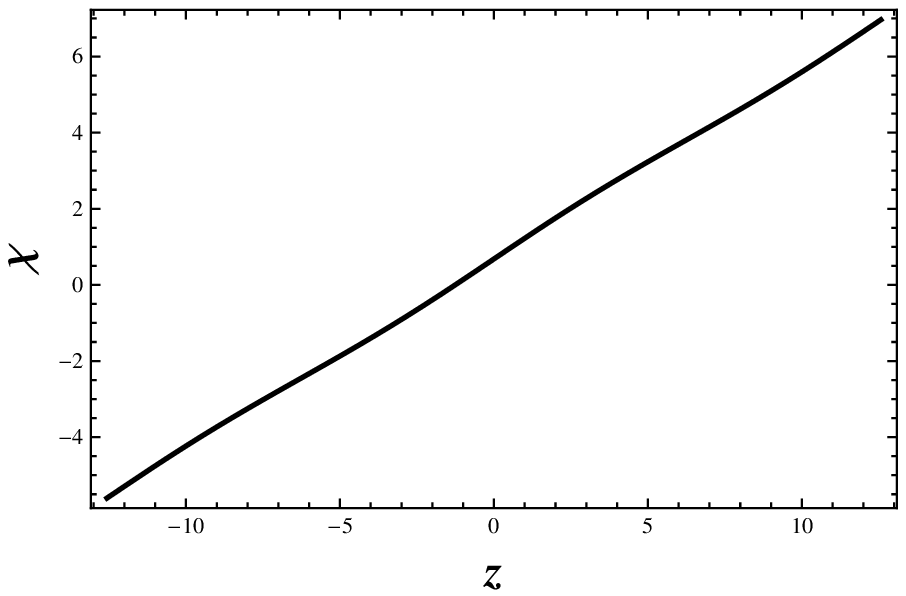} 
		      \hspace{.25cm}
		         \includegraphics[scale=0.35]{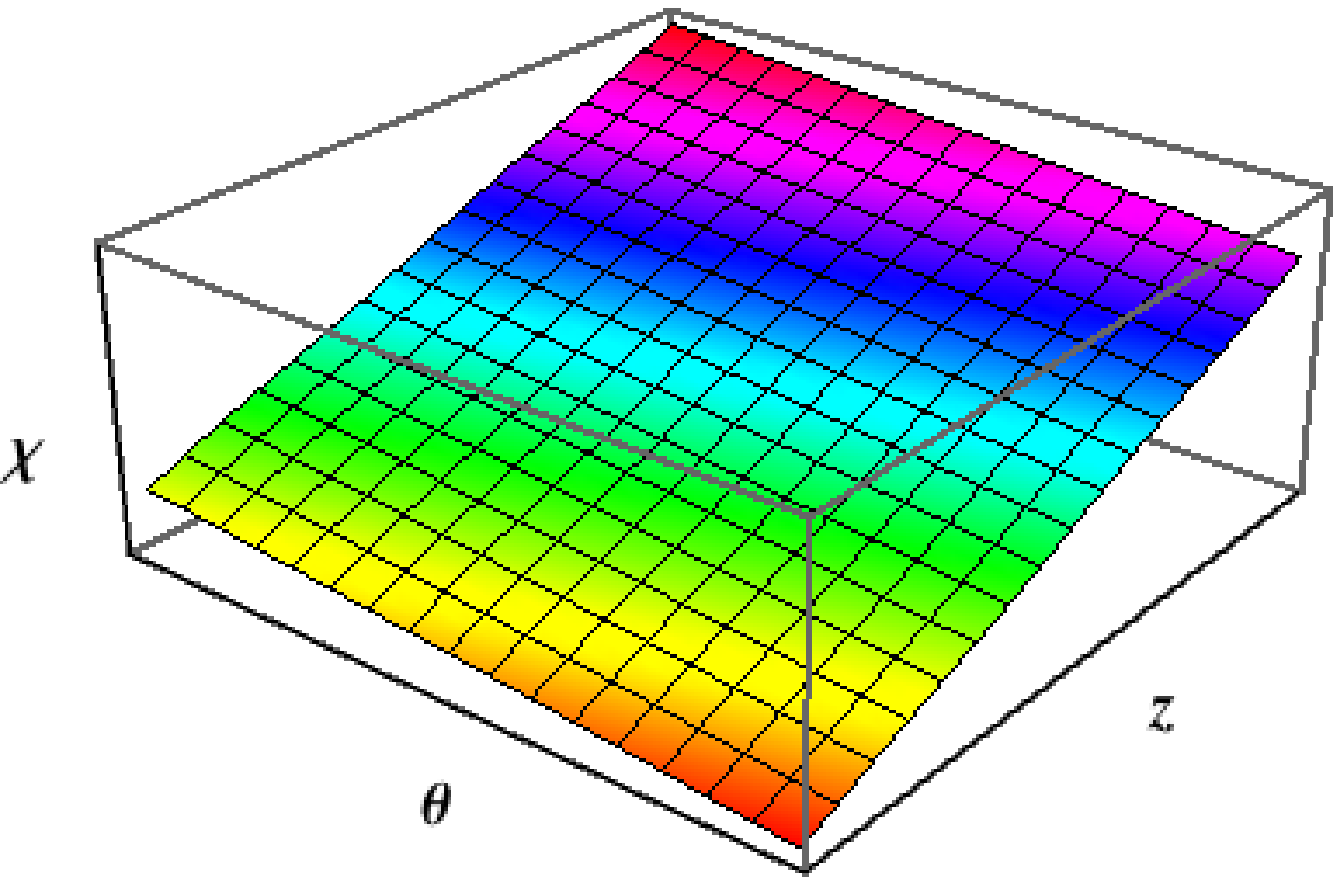} 
		         \hspace{.2cm}
		      \includegraphics[scale=0.45]{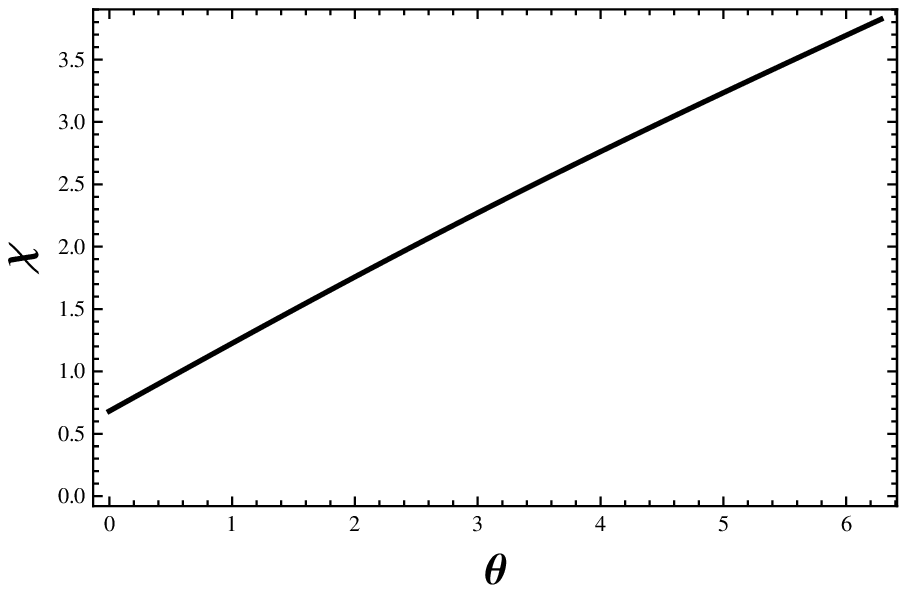} 
		   \caption{${\chi}$}
		   \label{fig:chi3D}
		\end{figure}	
			
		\section{$\mathbf{S}^{5}$ as a Hopf fibration over $\mathbf{P}^{2}$}
In this appendix we provide some details for realizing $\mathbf{S}^{5}$ as a Hopf fibration over $\mathbf{P}^{2}$,   
It is convenient to express the line element in terms of the following 1-forms first,
$$ \sigma_{1} = \frac{1}{2} (\text{$d\theta $} \cos (\psi )+\text{$d\phi $} \sin (\theta ) \sin (\psi ))$$
$$ \sigma_{2} = \frac{1}{2} (\text{$d\theta $} \sin (\psi )-\text{$d\phi $} \cos (\psi ) \sin (\theta )) $$
$$ \sigma_{3} = \frac{1}{2} (\text{$d\psi $}+\text{$d\phi $} \cos (\theta )) $$
In terms of these 1-forms, the metrics on $\mathbf{CP}^{2}$ and $\mathbf{S}^{5}$ may be written,
$$ ds^2_{\mathbf{CP}^{2}} = \text{$d\mu $}^2 +\sin ^2(\mu ) \left(\sigma_1^2+\sigma _2^2+\cos ^2(\mu ) \sigma _3^2\right) $$
$$ ds^2_{\mathbf{S}^{5}} = ds_{\mathbf{CP}^2}^2 +\left(d\varphi +\sin^2(\mu) \sigma _3 \right){}^2 $$
where $\varphi$ is the local coordinate on the Hopf fibre and $\mathbf{A} = \sin^2(\mu)\sigma _3={\sin^2(\mu)\over 2}(\text{$d\psi $}+\text{$d\phi $} \cos (\theta ))$ is the 1-form potential for the K\"ahler form on $\mathbf{CP}^2$. Now, the metric on $\mathbf{S}^5$ can be written as,
\be ds^2_{\mathbf{S}^5} = d\varphi^2 + d\mu^2 +  \sin^2\mu\left[\( d\psi + \cos \theta d\phi  \)  d\varphi +{1\over 4} \(d\theta^2 + d\phi^2 + d\psi^2 + 2 \cos\theta d\psi d\phi \)\right] \ee
		\newpage

	       \renewcommand{\baselinestretch}{1.05}

		\end{document}

We can write down a Bloch decomposition Note that the functions ${\cal Y}$ satisfy certain orthogonality relation as they are solutions of a Sturm-Liouville problem. So it is possible to specify boundary conditions on $H_{xy}$ such that the $\Delta = 4$ mode decouples from the other modes. It can be verfied that $H_{xy} = \Psi^{(AdS)}_{\Delta=4}\(w,p^{2}\)$ satisfies (\ref{waveequation}). The holographic two-point function can be obtained now by evaluating the second variation of the on-shell action of the fluctuations with respect to the boundary value of $H_{xy}$. The on-shell action is given by 
$$ S_{on-shell} =  \int d^{4} k dw d\theta e^{3C(w,\theta) + B(w,\theta)}e^{4w/L}\(\(\del_{w} \Psi^{AdS}_{\Delta =4}\)^{2} + e^{-2w/L}\del \Psi\(\Psi^{AdS}_{\Delta =4}\)^{2} \) $$
where we have used the fact that $\Psi^{AdS}_{\Delta =4}$ is independent of $\theta$. We can rewrite the on-shell action by expressing the periodic functions in the integrand as a Fourier series:
$$ S_{on-shell} = \mathop{\sum}_{\ell, n} \mathfrak{C}_{\ell n} \int d^{4} k dw d\theta e^{4w/L} e^{i \ell w/h + i n \theta}\(\(\del_{w} \Psi^{AdS}_{\Delta =4}\)^{2} + e^{-2w/L}k^{2}e^{4w/L}\(\Psi^{AdS}_{\Delta =4}\)^{2} \) + h.c. $$
		where $\mathfrak{C}_{\ell n}$ denotes the Fourier coefficients. The precise form of $\mathfrak{C}_{\ell n}$ can be determined, but it is not important for the present discussion. After integrating over $\theta$, we can see that only the $n = 0$ mode survives.